\documentclass[a4,11pt]{article}
\usepackage{amsmath}
\usepackage[top=2.54cm, bottom=2.54cm, left=2.54cm, right=1.50cm]{geometry}

\include{./Newcommands1}
\usepackage{graphicx}
\usepackage{float}
\usepackage{color}

\usepackage{hyperref}
\usepackage{natbib}
\usepackage{amssymb,amsmath,amsthm}

\usepackage{graphicx}
\usepackage{color}
\usepackage{fancyhdr}
\usepackage{marginnote}
\include{./Newcommands1}
\newcommand{\dsum}{\displaystyle\sum}
\newcommand{\dint}{\displaystyle\int}

\begin{document}
\vspace*{1cm}
\begin{center}
	
	 \textbf{{\large A General Bayesian Approach to Meet Different Inferential Goals in  Poverty Research for Small Areas}} \\ [14pt]
	
	{\fontsize{10pt}{1} \selectfont\textbf{Partha Lahiri$^{\text{1*}}$, and Jiraphan Suntornchost$^\text{2}$}} \\
	
	{\fontsize{9pt}{1} \selectfont\textbf{\textit{$^1$Department of Mathematics \& The Joint Program in Survey Methodology, University of Maryland, College Park,USA, Plahiri@umd.edu}}} \\
	{\fontsize{9pt}{1} \selectfont\textbf{\textit{$^2$Mathematics and Computer Science, Chulalongkorn University, Bangkok, 10330, Thailand, jiraphan.s@chula.ac.th
	}}} 
\end{center}
\vskip14pt
\begin{abstract}
Poverty mapping that displays spatial distribution of various poverty indices is most useful to policymakers and researchers when they are disaggregated into small geographic
units, such as cities, municipalities or other administrative partitions of a
country. Typically, national household surveys that contain welfare variables such as
income and expenditures provide limited or no data for small areas. It is well-known
that while direct survey-weighted estimates are quite reliable for national or large geographical
areas they are unreliable for small geographic areas. If the objective is to
find areas with extreme poverty, these direct estimates will often select small areas due
to the high variabilities in the estimates. Empirical best prediction and Bayesian
methods have been proposed to improve on the direct point estimates. However,
these estimates are not appropriate for different inferential purposes. For example, for
identifying areas with extreme poverty, these estimates would often select areas with large sample sizes.
In this paper, using databases used by the Chilean Ministry
for their Small Area Estimation production, we illustrate how appropriate
Bayesian methodology can be developed to address different inferential problems.
\end{abstract}
{\bf Keywords:} Bayesian model, Cross-Validation, Hierarchical models, Monte Carlo simulations

\section{Introduction}
\indent Eradication of poverty, one of the greatest challenges facing humanity,  has been the central tool to guide various public policy efforts in many countries. There has been remarkable progress in reducing extreme poverty rates, decreasing  by more than half since 1990. Despite this achievement, about $20\%$ of the world's population still live on less than 1.25 a day, and millions more are on the brink of poverty.  On September 25, 2015, the United Nations adopted the 2030 Agenda for Sustainable Development that compromises of 17 new Sustainable Development Goals (SDGs), beginning with a historical pledge to end poverty in all forms and dimensions by 2030 everywhere permanently;
see https://www.un.org/sustainabledevelopment/.  In order to achieve these goals, basic resources and services need to be more accessible to people living in vulnerable situations. Moreover, support for communities affected by conflict and climate related disasters needs to be raised. 

National estimate of an indicator usually hides important differences among different regions or areas with respect to that indicator. In almost all countries, these differences exist and can often be substantial.  The smaller the geographic regions for which indicators are available, the greater the effectiveness of interventions. Indeed this allows to reduce transfers to the non-poor and minimizes the risk that a poor person will be missed by the program. Ravallion (1994) found that Indian and Indonesian states or provinces are too heterogeneous for targeting to be effective. This underlines the need for production of estimates of indicators for small areas that are relatively homogenous.

It is now widely accepted that direct estimates of poverty based on household survey data are unreliable. There are now several papers available in the literature that attempt to improve on the direct estimates by borrowing strength from multiple relevant databases.  Hierarchical models that combine information from different databases are commonly used to achieve the goal because such models not only provide improved point estimates but also incorporate different sources of variabilities.  These models can be implemented using synthetic approach (e.g., Elbers et al. 2003), empirical best prediction approach (Fay and Herriott 1979; Franco and Bell 2015; Bell et al. 2016; Molina and Rao 2010; Casas-Cordero et al. 2016), and Bayesian approach (Molina et al. 2014).  See Jiang and Lahiri (2006), Pfeffermann (2013) and Rao and Molina (2015) for a review of different small area estimation techniques.

Empirical best prediction and hierarchical Bayesian methods (also shrinkage methods) have been employed in numerous settings, including studies of cancer incidence in Scotland (Clayton and Kaldor 1987), cancer mortality in France (Mollie and Richardson 1991),  stomach and bladder cancer mortality in Missouri cities (Tsutukawa et al. 1985), toxoplasmosis incidence in EI Salvador  (Efron and Morris 1975), infant mortality in New Zealand (Marshall 1991),  mortality rates for chronic obstructive pulmonary
disease (Nandram et al. 2000). The basic approach in all these applications is the same: a prior distribution of rates is posited and is combined with the observed rates to calculate the posterior, or stabilized, rates.

All the papers cited in the previous paragraph deal with the point estimation and the associated measure of uncertainty. However, in many cases, there could be different inferential goals where point estimates, whether empirical best prediction estimates or posterior means, though they can provide a solution, are not efficient.  For example, one inferential goal could be to flag a geographical area (e.g., municipality) for which the true poverty measure of interest exceeds a pre-specified standard. The point estimates can certainly flag such areas  but do not provide any reasonable uncertainty measure to assess the quality of such action.  It is not clear how to propose such a measure for a method based on direct estimates.  For the method based on posterior mean, one can perhaps propose a normal approximation using the posterior mean and posterior standard deviation to approximate the posterior probability of the true poverty measure exceeding the pre-specified standard.  In some cases, quality of such approximation could be questionable.  One can have a more complex inferential goal.  For example, we may be interested in identifying the worst geographical area with respect to the poverty measure.  In this case, the use of direct estimates for identification can be misleading when the sample sizes vary across the geographic units. The regions with small sample sizes will tend to have both high and low poverty indices merely because they have the largest variability. The method based on posterior means is not good either since it tends to identify areas with more samples; see Gelman and Price (1999).  Moreover, in either case there does not seem to be a clear way to produce a reasonable quality measure.

Gelman and Price (1999), Morris and Christiansen (1996), Langford and Lewis (1998), Jones and Spigelhalter (2011) discussed various inferential problems other than the point estimation.  For example, Morris and Christiansen (1996) outlined inferential procedures for identifying areas with extreme poverty.  Our approach is similar to theirs but applied for different complex parameters and used much more complex hierarchical model that is appropriate for survey data. 

We would like to stress that our proposed approach for solving different inferential problems  is fundamentally different from the constrained empirical /hierarchical Bayesian (Louis 1984; Ghosh 1992; Lahiri 1990)  and the triple-goal (Shen and Louis, 1998) approaches where the goal is to produce one set of estimates for different purposes.  In contrast, we propose to use the same synthetic data matrix generated from the posterior predictive distribution for different inferential purposes.  Other than this fundamental difference, our approach has a straightforward  natural way to produce appropriate quality measures.  In poverty research, Molina et al. (2014) suggested an interesting approach for estimating different poverty indices by generating a synthetic population from a posterior predictive density.  However, they restricted themselves to point estimates and their associated uncertainty measures and did not discuss how one would solve a variety of statistical inferences.
 
We outline a general approach to deal with different inferential goals and illustrate our methodology using the data used by the Chilean government for their small area poverty mapping system. Section 2 describes the Chilean data, the hierarchical model, a general poverty index, inferential approach to deal with various goals and data analysis.  In section 3, we provide some concluding remarks and direction for future research.

\section{Illustration of the proposed methodology using Chilean poverty data}
There is a consistent downward trend  in the official poverty rate estimates, which are the usual national survey-weighted direct estimates, in Chile since the early 90's. While this national trend is encouraging, there is an erratic time series trend in the direct estimates for small comunas (municipalities)  - the smallest territorial entity in Chile. Moreover, for a handful of extremely small comunas, survey estimates of poverty rates are unavailable for some or all time points simply because the survey design, which traditionally focuses on precise estimates for the nation and large geographical areas, excludes these comunas for some or all of the time points. In any case, direct survey estimates of poverty rates typically do not meet the desired precision for small comunas and thus the assessment of implemented policies is not straightforward at the comuna level. In order to successfully monitor trends, identify influential factors, develop effective public policies and eradicate poverty at the comuna level, there is a growing need to improve on the methodology for estimating poverty rates at this level of geography. In this section, we use the Chilean case to illustrate our Bayesian approach to answer a variety of research questions.

\subsection{Data used in the Analysis}
 To illustrate our Bayesian approach, we use a household survey data  as the primary source of information and comuna level summary statistics obtained from different administrative data sources as supplementary sources of information.  We now provide a brief description of the primary and supplementary databases.  Further details can be obtained from Casas-Cordero et al. (2016).

\subsubsection{The Primary Data Source: The CASEN 2009 Data}
The Ministry of Social Development estimates the official poverty rates using the National Socioeconomic Characterization Survey, commonly referred to as the  CASEN. The Ministry has been conducting CASEN since 1987 every two or three years.  The CASEN is a household survey collecting a variety of information of Chilean households and persons, including information about income, work, health, subsidies, housing and others. The Ministry calculates poverty rate estimates at national, regional and municipality (comuna) levels.  The Ministry is the authority specified by the Chilean law to deliver poverty estimates for all the 345 comunas in Chile. These estimates are used, along with other variables, to allocate public funding to municipalities. In a joint effort by the Ministry and United Nations Development Programme (UNDP), a Small Area Estimation (SAE) official system was developed for estimating poverty rates at comuna level using the CASEN 2009 survey.  The Chilean method is based on an empirical Bayesian method using an area level Fay-Herriot Model (Fay and Herriott 1979) to combine the CASEN survey data with a number of administrative databases.  The SAE system provides point estimates and parametric bootstrap confidence intervals (see, e.g., Chatterjee et al. 2008; Li and Lahiri 2010) for the Chilean comunas.

The CASEN 2009 used a stratified multistage complex sample of approximately $75,000$ housing units from $4,156$ sample areas. The entire Chile was divided into a large number of sections (Primary Stage Units - PSUs). The PSUs were then grouped into strata on the basis of two geographic characteristics:   comuna and urban/rural classification. Overall, there were 602 strata in the CASEN 2009 survey and multiple PSUs were sampled per stratum. The probability of selection for each PSU in a stratum was proportional to the number of housing units in the (most recently updated) 2002 Census file.

Prior to the second stage of sampling, listers were sent to the sampled PSUs to update the count of housing units.
 This procedure was implemented in both urban and rural areas.
 In the second stage of sampling, a sample of housing units was selected within the sampled PSUs.
 The probability of selection for each housing unit (Secondary Stage Unit--SSU) is the same within each PSU.
 On the average, 16-22 housing units were selected within each PSU by implementing a procedure
 that used a random start and a systematic interval to select the units to be included in the sample.

\subsubsection{Administrative Data at the Comuna Level}
Casas-Cordero et al. (2016) carried out an extensive task to identify a set of auxiliary variables derived from different administrative records of different agencies. In this paper, we use the same set of auxiliary variables for the comunas we selected for illustrating our approach.  For completeness, we list them below:
\begin{description}
\item (1) Average wage of workers who are not self-employed,
\item (2) Average of the poverty rates from CASEN 2000, 2003, and 2006,
\item (3) Percentage of population in rural areas,
\item (4) Percentage of illiterate population,
\item (5) Percentage of population attending school.
\end{description}
Like in Casas-Cordero et al. (2016), we also use arcsin square-root transformation for all the auxiliary variables except the first one for which we use logarithmic transformation.  We note that our approach is general and can use a different set of auxiliary variables that may be deemed appropriate in the future.
\subsection{The Foster-Greer-Thorbecke (FGT) poverty indices }
Currently, the Chilean government publishes headcount ratios or poverty rates for the nation and its comunas.  To present our approach in a general setting, we consider a general class of poverty indices commonly referred to as the FGT indices, after the names of the three authors ( Foster et al. 1984). To describe the FGT index, we first introduce the following notations:
\begin{description}
\item $N_c$: the total number of households in comuna $c$, 
\item $U_c$: the number of urbanicity statuses for comuna $c$; since for urbanicity status, we use urban and rural statuses only, $U_c$ is either $1$ or $2$ for a given comuna,

\item $k_{u}$: the fixed poverty line for urban-rural classification $u$ ($u=1$ and $u=2$ for urban and rural, respectively); a fixed poverty line determined nationally that is constant for all urban residents and another for all rural resident,

\item $M_{cu}$: the total number of PSUs in the universe for  urban-rural classification  $u$ of comuna $c$,
\item $N_{cup}$: the total number of households in the universe of the  
PSU $p$ belonging to the  urban-rural classification $u$ of comuna $c$,
\item $y_{cuph}$: the per-capita income of household $h$ (that is, total income of the household divided by the number of household members) in  PSU $p$, urban-rural classification $u$, and comuna $c$.
\end{description}
In our context, the class of FGT indices for comuna $c$ is given by

\[{Q}_{c;\alpha}=\dfrac{1}{N_{c}}\displaystyle\sum_{u=1}^{U_c}\displaystyle\sum_{p=1}^{M_{cu}}
\displaystyle\sum_{h=1}^{N_{cup}}  g_{\alpha}(y_{cuph}),\]
where
$g_{\alpha}(y_{cuph}) =\left(\dfrac{k_{u}-y_{cuph}}{k_{u}}\right)^{\alpha}\mathcal{I}(y_{cuph} < k_{u})$,
$\alpha$ is a \lq\lq sensitivity\rq\rq  parameter ($\alpha =0, 1, 2$ corresponding to poverty ratio, poverty gap, and poverty severity, respectively).

\subsection{Hierarchical Model}
A hierarchical model could be effective in capturing different salient features of the CASEN survey data and in linking comuna level auxiliary variables derived from different administrative records.  We consider the following working hierarchical model to illustrate our general approach for inference.  We call the model a working model because we recognize that it is possible to improve on it in the future.  But this model will suffice to illustrate the central theme of the paper, which is how to carry out a particular inferential procedure given a hierarchical model.
 Let $T_{cuph}=T(y_{cuph})$ be a given transformation on the study variable $y_{cuph}$, which we can take before application of the following hierarchical model.  For the application of this paper, we consider $T(y_{cuph})=\ln (y_{cuph}+1).$
 We consider the following hierarchical model for the sampled units:


$\begin{array}{lrll}
\mbox{Level 1:}&T_{cuph}|\theta_{cup}, \sigma_{T} & \stackrel{ind}\sim&  N\big(\theta_{cup}, \sigma^2_{T} \big),\notag
\\ \mbox{Level 2:}&\theta_{cup}|\mu_{cu},\sigma_\theta &\stackrel{ind}\sim &N\big( \mu_{cu},\sigma^{2}_{\theta} \big),\label{eq:model}
\\  \mbox{Level 3:}&\mu_{cu}|\beta_{u},\sigma_{\mu} &\stackrel{ind}\sim& N\big( \mathbf{x}_{c}^{T} \beta_{u},\sigma_{\mu}^2 \big).\notag
\end{array}$

%
%

We follow the recommendation of Gelman (2015) in assuming weakly informative priors for the regression coefficients $\beta$ and the variance components.  For example, we assume independent $N(0,1)$ prior for all regression coefficients and independent half normal prior for the standard deviations.

\subsection{Inferential Approach}
We first note that the inference on $Q_{c,\alpha}$ is equivalent to that of
\[Q_{c;\alpha} =\frac{1}{N_c}\dsum_{u=1}^{U_c}\dsum_{p=1}^{M_{cu}}\dsum_{h=1}^{N_{cup}}  g_{\alpha}\left( T^{-1}(T_{cuph})\right),\]
where $T$ is a monotonic function (e.g., logarithm). Under full specification of the model for the finite population, one can make inferences about $Q_{c;\alpha}$ in a standard way.  However, full specification of model for the unobserved units of the finite population seems to be a challenging task.  To this end, appealing to the law of large numbers, we first approximate $Q_{c;\alpha}$ by $\tilde{Q}_{c;\alpha}^{P}$, where 
$$\tilde{Q}_{c;\alpha}^{P}=\frac{1}{N_c}\sum_{u=1}^{U_c}\sum_{p=1}^{M_{cu}}\sum_{h=1}^{N_{cup}}
{\rm} E\left\{g_{\alpha}\left (T^{-1}(T_{cuph})\right )|\theta_{cup},\sigma_T\right\}.$$  This is reasonable under Level 1 of the hierarchical model (even without the normality assumption) since  $N_c$ is typically large.  We then propose the following approximation to $\tilde{Q}_{c;\alpha}^{P}$.   
\begin{equation}
\tilde {Q}_{c;\alpha}\equiv \tilde {Q}_{c;\alpha}(\theta_{c},\sigma_{T})=\dsum_{u=1}^{U_c}\dsum_{p=1}^{m_{cu}}\dsum_{h=1}^{n_{cup}} w_{cuph} \mathbb{E}\big\{g_{\alpha}\left( T^{-1}(T_{cuph})\right) | \theta_{cup},\sigma_{T} \big\},\label{eq:qcal}
\end{equation}
where
\begin{description}
\item $w_{cuph}$ is the survey weight for the household $h$ in the PSU $p$ within urbanicity $u$ in comuna $c$,
\item $\theta_{c}=\mbox{col}_{u,p} \theta_{cup}$; a $\displaystyle\sum_{u=1}^{U_c}m_{cu} \times 1$ column vector (we follow the notation of Prasad and Rao 1990),
\item $g_{\alpha}\left( T^{-1}(T_{cuph})\right)=\left \{\dfrac{k_{u}-\left( T^{-1}(T_{cuph})\right)}{k_{u}}\right \}^{\alpha}I\big(T_{cuph} \leq l_{u} \big)$,
\item $l_{u}=\ln (k_u+1)$, the poverty line of the urbanicity $u$ in the transformed scale.
\end{description} 
The weights are scaled within each comuna so that the sum of the weights for all households equals 1. 
In the last approximation,  we assume that the scaled survey weight $w_{cuph}$ represents proportion of units in the finite population (including the unit $cuph$) of comuna $c$  that are similar to the unit $cuph$.

The calculations of (\ref{eq:qcal}) under the model described in section \ref{eq:model} for the cases where $\alpha=0$ and $\alpha=1$ can be done through the following formula:
\\ For $\alpha=0$,
\begin{align*}
\mathbb{E}\big\{g_{0}(\left( T^{-1}(T_{cuph})\right) | \theta_{cup},\sigma_{T}^{2} \big\}&=\dint_{\dfrac{-\theta_{cup}}{\sigma_{T}}}^{\dfrac{l_{u}-\theta_{cup}}{\sigma_{T}}}\phi(z | \theta_{cup},\sigma_{T}^{2})dz\notag
\\&=\Phi\Big( \dfrac{l_{u}-\theta_{cup}}{\sigma_{T}}\Big)-\Phi\Big( \dfrac{-\theta_{cup}}{\sigma_{T}}\Big),\label{eq:alpha0}
\end{align*}
where $\phi$ and $\Phi$ are the density function and the distribution function of the standard normal distribution, respectively.
\\ For $\alpha=1$,
\begin{align*}
\mathbb{E}\big\{g_{1}(T^{-1}(T_{cuph})) | \theta_{cup},\sigma^{2} \big\}&=E \left\{\left( \dfrac{\exp(l_{u})-\exp(T_{cuph})}{k_{u}}  \right)I\big( T_{cuph} \leq l_{u} \big)\right\}
\\&=\dfrac{1}{k_{u}} \displaystyle\int_{-\dfrac{\theta_{cup}}{\sigma_{T}}}^{\dfrac{l_{u}-\theta_{cup}}{\sigma_{T}}} \left( \exp(l_{u})-\exp(\sigma_{T}z+\theta_{cup})\right)\phi(z|\theta_{cup,\sigma^2})dz
\\&= \dfrac{\exp(l_{u})}{k_{u}} \Big[\Phi\Big( \dfrac{l_{u}-\theta_{cup}}{\sigma_{T}}\Big)-\Phi\Big( \dfrac{-\theta_{cup}}{\sigma_{T}}\Big)  \Big]
\\&~~~-\dfrac{\exp(\theta_{cup}+\dfrac{\sigma^2_{T}}{2})}{k_{u}}\left[ \Phi\Big( \dfrac{l_{u}-\theta_{cup}-\sigma^2_{T}}{\sigma_{T}}\Big)-\Phi\Big( \dfrac{-\theta_{cup}-\sigma^2_{T}}{\sigma_{T}}\Big)     \right].
\end{align*}
where we use the fact that $\displaystyle\int_{a}^{b}\exp(\sigma z)\phi(z)dz=\exp(\dfrac{\sigma^2}{2})\Big[ \Phi(b-\sigma)-\Phi(a-\sigma)  \Big]$ to obtain the last equation.
\vskip .2in

In order to carry out a variety of inferential problems about $\tilde {Q}_{c;\alpha}$ for
a given $\alpha$, we use the Monte Carlo Markov Chain (MCMC). The procedures are described below.
\\Let $C$ be the number of comunas covered by the model and $R$ be the number of MCMC samples after burn-in.
Let $\theta_{c;r}$ and $\sigma_{T;r}$ denote the $r$th MCMC draw of $\theta_c$ and $\sigma_T$, respectively ($r=1,\ldots,R)$.
We define the $C\times R$, matrix $\tilde {Q}_{\alpha}^s=(\tilde {Q}_{(c,r);\alpha}^s),$ where the $(c,r)$ entry is defined as
$$\tilde {Q}_{(c,r);\alpha}^s\equiv \tilde {Q}_{c;\alpha}^s(\theta_{c;r},\sigma_{T;r}).$$  
This matrix $\tilde {Q}_{\alpha}^s$ provides samples generated from the posterior distribution of $\{\tilde Q_{c,\alpha},\; c=1,\ldots,C\}$  and so is adequate for solving a variety of inferential problems in a Bayesian way.  We now elaborate on the following three different inferential problems:

\begin{description}
\item (1) Point estimation of an indicator of interest and the associated measure of uncertainty: This is the focus of current poverty mapping research in both classical and Bayesian approaches.  Under squared error loss function, the Bayes estimate of $Q_{c;\alpha}$  for comuna $c$ and the associated measure of uncertainty are the posterior mean and posterior standard deviation of $\tilde {Q}_{c;\alpha}\equiv \tilde {Q}_{c;\alpha}(\theta_{c},\sigma_{T})$, respectively.  These  can be approximated by the average and standard deviation across columns of $\tilde {Q}_{\alpha}^s$, respectively, for the row $c$, which  corresponds to the comuna $c$.

\item (2) Identification of comunas that are not in conformity with a given standard of a poverty indicator: In this inferential problem, the goal is to flag a comuna for which the true poverty indicator (e.g., poverty rate) exceeds a pre-specified standard, say $a$.  In this case, point estimates, whether direct estimates or posterior means, do not give any idea about the quality of flagging a comuna that does not meet the given standard.  A reasonable Bayesian solution for this inferential problem is to flag comuna $c$ for not meeting the given standard if  the posterior probability $P(\tilde Q_{c;\alpha} >a|\mbox{data})$ is greater than a specified cutoff, say $0.5$.  This posterior probability for comuna $c$ can be easily approximated by the proportion of columns of $\tilde Q_{c;\alpha} ^s$ exceeding the threshold for row $c$.  If the posterior distribution of $\tilde Q_{c;\alpha} $ is approximately normal, then one can alternatively use the posterior mean and posterior standard deviation to approximate the posterior probability.  However, such an approximation may not perform well in many situations.

\item (3) Identification of the worst (best) comuna, i.e., the comuna with the maximum (minimum) value of the poverty indicator:  A common solution is to identify the comuna with the maximum (minimum) point estimate of the indicator.  Evidently, the use of direct point estimates would be quite misleading since such a method may identify a small comuna as being the worst (best) in terms of the indicator, even though it is not, simply because of high variability in the direct estimates.  The Bayesian point estimates (posterior means) are definitely better than the direct estimates as they have generally less variability.  However, the use of posterior means alone does not provide any quality measure associated with the identification of the worst (best) comuna.  Even the use of posterior means along with posterior standard deviations does not help either as posterior standard deviations relate to the individual areas.  A reasonable Bayesian solution in this case would be to compare the posterior probabilities $P(\tilde Q_{c;\alpha} \ge \tilde Q_{k;\alpha} \;\forall k|\mbox{data})$ for different comunas and select the worst (best) comuna for which this posterior probability is the maximum (minimum).  Thus, along with the  identification of the worst (best) comuna, we also obtain these posterior probabilities suggesting a quality of the identification of worst (best) comuna.  We can use $\tilde Q_{\alpha}^s$ matrix to approximate these posterior probabilities. For row $c$ and column $r$ of $\tilde Q_{\alpha}^s$ corresponding to comuna $c$ and MCMC replicate $r$, respectively, we can create a binary variable indicating if the comuna is the worst (best) among all comunas.  The posterior probability for this comuna $P(\tilde Q_{c;\alpha} \ge \tilde Q_{k;\alpha} \;\forall k|\mbox{data})$ can then be approximated by the average of these binary observations across $R$ columns.

\end{description}


\subsection{Numerical Results}
As mentioned in the introduction, a number of researchers focused on the problem of estimation and its measure on uncertainty.  While our general approach can address this problem, we choose to illustrate the  general Bayesian approach for the relatively understudied inferential problems related to the identification of ares with extreme poverty (e.g., the  second and the third inferential problems mentioned in section 2.4).  The data analysis presented in this section is based on the hierarchical model stated in section 2.3 implemented on  CASEN 2009 data for a given region containing 15 comunas and comuna level auxiliary variables listed in section 2.1  We illustrate our methodology for poverty rates ($\alpha=0$) and poverty gaps ($\alpha=1$), two important poverty measures in the FGT class of poverty indices. After 10,000 burn-in, we generate $15\times 10000$ matrix $\tilde Q_{c;\alpha} ^s$ for $\alpha=0$ (corresponding to poverty rate index) and $\alpha=1$ (poverty gap index). We checked the convergence of MCMC convergence using the potential scale reduction factor introduced by Gelman and Rubin (1992). 

Table 1 addresses the second inferential goal, i.e., flagging the comunas that do not meet certain pre-specified standard for poverty rate.  Table 2 is similar to Table 1 except that this is for the poverty gap measure.  We use three different standards based on three different multipliers ($1.10,1.25$ and $1.50$) of the regional direct estimate of the respective measure.  These standards are for illustration only and our approach can use any other standards that are deemed reasonable.  We need a cutoff for these posterior probabilities in order to flag comunas that do not meet the given standard. To illustrate our approach, we use 0.5 as the cutoff.  In other words, a comuna is deemed out of the range with respect to the pre-specified standard if the posterior probability is more than 0.5.  Comunas 33 and 13 do not meet all three standards for both poverty rate and poverty gap measures.  Other comunas meet the more liberal standard (1.5 times the regional poverty measure) with respect to both poverty rate and poverty gap measures.  In contrast, when the standard is very conservative (1.1 times the regional poverty measure) all the comunas are not in conformity with the given standard.  For a moderate standard (1.25 times the regional poverty measure), comunas 33, 13, 22, 18, 2, 6, 45, 16, 30 do not satisfy the standard in terms of poverty rate measure.  The comunas 21, 5, 17 and 15 are added to the list when we consider the poverty gap measure.  The standard and the cut-off to be used are subjective, but the Bayesian approach with different standard and cutoff combinations should give policy makers some useful guidance in making certain policy decisions.

Table 3 displays approximations (by MCMC) to the posterior probabilities of a comuna being the worst (Prob.Max) as well as the best (Prob.Min) in terms of both poverty rate and poverty gap measures.  According to the Prob.Max criterion, comuna 33 stands out as the worst comuna in terms of both poverty rate and poverty gap measures.  According to Prob.min criterion, comuna 8 emerges as the best comuna in terms of both poverty rate and poverty gap measures.  These probabilities are also giving us a good sense of the confidence we can place on our decision, which is not possible with poverty rate and poverty gap estimates alone.

\begin{table}
\caption{The posterior probabilities that poverty rate for a comuna exceeds three \newline different
 thresholds; $Q_{r,0}$ is direct estimate of regional poverty rate.}
\centering
\begin{tabular}{|r|r|r|r|}
  \hline&&&\\
 & $P(\widetilde{Q}_{c;0}>1.10{Q}_{r,0}|\mbox{data} )$ &  $P(\widetilde Q_{c;0} >1.25{Q}_{r,0}|\mbox{data} )$  &  $P(\widetilde Q_{c; 0} >1.50{Q}_{r,0} |\mbox{data})$ \\&&&\\
  \hline&&&\\
33 & 1.0000 & 0.9995 & 0.6172 \\
13 & 1.0000 & 0.9988 & 0.5636 \\
22 & 0.9952 & 0.7962 & 0.0314 \\
18 & 0.9904 & 0.6996 & 0.0100 \\
2 & 0.9834 & 0.4939 & 0.0005 \\
6 & 0.9809 & 0.5331 & 0.0006 \\
45 & 0.9786 & 0.5755 & 0.0032 \\
16 & 0.9721 & 0.5157 & 0.0015 \\
30 & 0.9662 & 0.5086 & 0.0024 \\
21 & 0.9362 & 0.3925 & 0.0013 \\
5 & 0.9356 & 0.3878 & 0.0010 \\
17 & 0.9258 & 0.3840 & 0.0012 \\
15 & 0.9185 & 0.3643 & 0.0015 \\
25 & 0.8822 & 0.2524 & 0.0002 \\
43 & 0.8755 & 0.2266 & 0.0000 \\
38 & 0.8612 & 0.2209 & 0.0003 \\
27 & 0.8466 & 0.2139 & 0.0002 \\
26 & 0.8425 & 0.3259 & 0.0022 \\
51 & 0.8365 & 0.2941 & 0.0009 \\
24 & 0.7835 & 0.1216 & 0.0000 \\
29 & 0.7111 & 0.0995 & 0.0000 \\
28 & 0.7030 & 0.1085 & 0.0000 \\
31 & 0.6771 & 0.0700 & 0.0000 \\
35 & 0.6694 & 0.1018 & 0.0000 \\
36 & 0.6404 & 0.0731 & 0.0000 \\
41 & 0.6142 & 0.0591 & 0.0001 \\
37 & 0.6041 & 0.0775 & 0.0000 \\
7 & 0.5705 & 0.0386 & 0.0000 \\
47 & 0.5179 & 0.0417 & 0.0000 \\
\hline
\end{tabular}
\end{table}
	
		\begin{table}
\caption{Posterior probabilities that poverty gap for a given comuna exceeds three \newline different
 thresholds; $Q_{r,1}$ is direct estimate of regional poverty gap.}

\centering
\begin{tabular}{|r|r|r|r|}
 \hline &&&\\
 & $P(\widetilde{Q}_{c;1}>1.10{Q}_{r,1}|\mbox{data} )$ &  $P(\widetilde{Q}_{c;1}>1.25{Q}_{r,1}|\mbox{data} )$  &  $P(\widetilde{Q}_{c;1}>1.50{Q}_{r,1} |\mbox{data})$ \\&&&\\
  \hline&&&\\
33 & 1.0000 & 0.9998 & 0.9266 \\
13 & 1.0000 & 0.9994 & 0.9060 \\
22 & 0.9966 & 0.9143 & 0.2635 \\
18 & 0.9918 & 0.8327 & 0.1195 \\
2 & 0.9893 & 0.7516 & 0.0395 \\
45 & 0.9827 & 0.7577 & 0.0781 \\
6 & 0.9824 & 0.7174 & 0.0300 \\
16 & 0.9792 & 0.7189 & 0.0490 \\
30 & 0.9693 & 0.6764 & 0.0489 \\
21 & 0.9467 & 0.5871 & 0.0320 \\
5 & 0.9420 & 0.5656 & 0.0240 \\
17 & 0.9339 & 0.5592 & 0.0292 \\
15 & 0.9333 & 0.5631 & 0.0337 \\
38 & 0.9001 & 0.4329 & 0.0081 \\
25 & 0.8923 & 0.4203 & 0.0079 \\
43 & 0.8802 & 0.3812 & 0.0070 \\
26 & 0.8751 & 0.5310 & 0.0657 \\
27 & 0.8745 & 0.3970 & 0.0104 \\
51 & 0.8540 & 0.4497 & 0.0305 \\
24 & 0.8223 & 0.2674 & 0.0018 \\
29 & 0.7700 & 0.2401 & 0.0026 \\
28 & 0.7441 & 0.2390 & 0.0026 \\
31 & 0.7321 & 0.1848 & 0.0005 \\
35 & 0.6924 & 0.2091 & 0.0026 \\
36 & 0.6671 & 0.1631 & 0.0006 \\
37 & 0.6399 & 0.1772 & 0.0021 \\
7 & 0.6376 & 0.1243 & 0.0002 \\
 41 & 0.6355 & 0.1365 & 0.0003 \\
 47 & 0.5586 & 0.1095 & 0.0003 \\ \hline
\end{tabular}
\end{table}

\begin{table}[H]
\caption{Posterior probability that poverty rate or poverty gap for a given comuna is the maximum (Prob.Max) or the minimum (Prob.Min)}

\centering
{\small\begin{tabular}{|c|r|r|r|r|}
  \hline
\\ COMUNA & \multicolumn{2}{ |c| }{Poverty Rate}&\multicolumn{2}{ |c| }{Poverty Gap}
\\\cline{2-5}\\	
 & Prob.Max & Prob.Min & Prob.Max.Gap & Prob.Min.Gap \\	
  \hline
33 & 0.5126 & 0.0000 & 0.5246 & 0.0000 \\
   13 & 0.4496 & 0.0000 & 0.4301 & 0.0000 \\
   22 & 0.0169 & 0.0000 & 0.0215 & 0.0000 \\
   18 & 0.0051 & 0.0000 & 0.0044 & 0.0000 \\
   45 & 0.0025 & 0.0000 & 0.0031 & 0.0000 \\
   17 & 0.0021 & 0.0000 & 0.0021 & 0.0000 \\
   26 & 0.0021 & 0.0000 & 0.0042 & 0.0000 \\
   30 & 0.0017 & 0.0000 & 0.0017 & 0.0000 \\
   21 & 0.0013 & 0.0000 & 0.0016 & 0.0000 \\
   15 & 0.0010 & 0.0000 & 0.0011 & 0.0000 \\
   16 & 0.0009 & 0.0000 & 0.0012 & 0.0000 \\
   51 & 0.0008 & 0.0000 & 0.0009 & 0.0000 \\
   6 & 0.0007 & 0.0000 & 0.0006 & 0.0000 \\
   5 & 0.0006 & 0.0000 & 0.0006 & 0.0000 \\
   2 & 0.0005 & 0.0000 & 0.0009 & 0.0000 \\
   27 & 0.0005 & 0.0000 & 0.0004 & 0.0000 \\
   38 & 0.0005 & 0.0000 & 0.0006 & 0.0000 \\
   25 & 0.0003 & 0.0000 & 0.0001 & 0.0000 \\
   35 & 0.0001 & 0.0000 & 0.0002 & 0.0000 \\
   41 & 0.0001 & 0.0000 & 0.0000 & 0.0000 \\
   43 & 0.0001 & 0.0000 & 0.0000 & 0.0000 \\
   28 & 0.0000 & 0.0000 & 0.0001 & 0.0000 \\
   40 & 0.0000 & 0.0001 & 0.0000 & 0.0001 \\
   46 & 0.0000 & 0.0001 & 0.0000 & 0.0002 \\
   44 & 0.0000 & 0.0002 & 0.0000 & 0.0004 \\
   23 & 0.0000 & 0.0008 & 0.0000 & 0.0012 \\
   10 & 0.0000 & 0.0009 & 0.0000 & 0.0005 \\
   14 & 0.0000 & 0.0009 & 0.0000 & 0.0011 \\
   3 & 0.0000 & 0.0052 & 0.0000 & 0.0047 \\
   34 & 0.0000 & 0.0057 & 0.0000 & 0.0079 \\
   4 & 0.0000 & 0.0075 & 0.0000 & 0.0089 \\
   12 & 0.0000 & 0.0121 & 0.0000 & 0.0139 \\
   48 & 0.0000 & 0.0186 & 0.0000 & 0.0237 \\
   42 & 0.0000 & 0.0240 & 0.0000 & 0.0268 \\
   1 & 0.0000 & 0.3929 & 0.0000 & 0.3945 \\
   8 & 0.0000 & 0.5310 & 0.0000 & 0.5161 \\
   \hline
\end{tabular}}
\end{table}

\section{Concluding Remarks}
We try to bring awareness of inappropriateness of using point estimates for all inferential purposes and propose a general  Bayesian approach to solve different inferential problems in the context poverty mapping.  The proposed approach not only provides an action relevant to the inferential problem but also provides a way to assess the quality of such action.  To make the methodology user-friendly one can store the $Q_{\alpha}^s$ matrix of size $C\times R$, where $C$ is the number of comunas and $R$ is the number of MCMC replications.  This way the users do not need to know how to generate this matrix, which requires knowledge of advanced Bayesian computing.  Once the user has access to this generated matrix, he/she can easily carry out a variety of statistical analysis such as the ones presented in the paper with greater ease. While we illustrate the approach for FGT poverty indices the approach is general and can deal with other important indices such as the ones given in sustainable development goals.  We have taken one working model to illustrate the approach, but the approach is general and can be applied to other models that are deemed appropriate in other projects.


\end{document}